\documentclass{PoS}
\usepackage{graphicx}
\usepackage{lineno}
\input{babarsym.tex}

\title{Light Higgs and Dark Photon Searches at {\babar}}

\ShortTitle{Light Higgs and Dark Photon Searches at {\babar}}

\author{\speaker{Alejandro P\'erez P\'erez}\\
        INFN - sezione di Pisa\\
        E-mail: \email{luis.alejandro.perez@pi.infn.it}\\
        {\it On behalf of the {\babar} Collaboration}}

\abstract{Several new-physics (NP) models predict the existence of low-mass Higgs states and light dark matter candidates.
Previous {\babar} searches have given null results for these new states and have excluded large regions of the NP models 
parameter space. We report on new searches on light Higgs and light dark matter at {\babar} using the $516~{\rm fb}^{-1}$ 
of data collected with the {\babar} detector at the PEP-II asymmetric-energy $e^+e^-$ collider at the SLAC National 
Accelerator Laboratory.}

\FullConference{XXI International Workshop on Deep-Inelastic Scattering and Related Subject -DIS2013,\\
		22-26 April 2013\\
		Marseilles,France}

\begin{document}
\section{Light Higgs searches at {\babar}}

According to the most accepted theories, the fundamental particles acquire mass through the Higgs 
mechanism~\cite{Higgs_mechanism}, which requires the existence of at least one scalar state
called the Higgs boson. In the standard model (SM) of particle physics~\cite{SM} there is 
only a single Higgs boson, and present experimental evidence by the ATLAS and CMS collaborations at LHC 
suggest a Higgs-like particle with a mass around $126~{\rm GeV/c^2}$~\cite{HiggsLike_discovery}. The 
Minimal Super-Symmetric SM (MSSM) solves the hierarchy problem of the SM by extending the 
Higgs sector, the masses of the Higgs bosons depending on a parameter $\mu$~\cite{MSSM}. 
The MSSM fails to explain why the value of the $\mu$ is of the order of the electroweak scale, 
many order of magnitude below the natural Plank scale. The next-to-minimal super-symmetry SM 
(NMSSM)~\cite{NMSSM} solves this so-called "naturalness problem" by adding a singlet chiral super-field 
to the MSSM. As a result the NMSSM contains two charged Higgs bosons, three neutral CP-even bosons, and 
two CP-odd bosons. The lightest CP-odd state, $A^0$, could have a mass smaller that twice the 
$b$-quark~\cite{NMSSM}, escaping detection at LEP, but making it detectable via 
$\Upsilon(nS)\rightarrow\gamma A^0$ decays at the B-factories ({\babar} and Belle)~\cite{LightHiggsRadiative}.
The branching fraction of $\Upsilon(nS)\rightarrow \gamma A^0$ could be as large as $10^{-4}$ depending of the 
values of the couplings~\cite{NMSSM}. All this makes the B-factories experimental environment 
an ideal place to search for light Higgs bosons. The expected width of this $A^0$ is expected to be 
smaller than the current experimental resolution on its mass, so its width is always 
neglected in the searches performed up to date.

Previous {\babar} searches for $A^0$ production in several final states have given null results, 
including $\Upsilon(2S,3S)\rightarrow\gamma A^0(\rightarrow \mu^+\mu^-,~\tau^+\tau^-)$ and 
$\Upsilon(1S)\rightarrow\gamma A^0(\rightarrow invisible)$~\cite{BaBar_previous}. Similar searches 
have been done by CLEO in the di-$\mu$ and di-$\tau$ with $\Upsilon(1S)\rightarrow\gamma A^0$ 
decays~\cite{CLEO_light_higgs}, and more recently by BESIII in $J/\psi\rightarrow \gamma A^0(\rightarrow \mu^+\mu^-)$
~\cite{BESIII_light_higgs}, and by CMS in $pp \rightarrow A^0 (\rightarrow \mu^+\mu^-)$~\cite{CMS_light_higgs}.

Two datasets of the {\babar} experiment can be used to search for the $A^0$ in the radiative decays 
$\Upsilon(nS)\rightarrow\gamma A^0$ of the narrow $\Upsilon(nS)$ resonances, with integrated luminosities 
of $27.9~{\rm fb}^{-1}$ at the centre-of-mass (CM) energy of the $\Upsilon(3S)$ 
and $13.6~{\rm fb}^{-1}$ at the $\Upsilon(2S)$, they contain $N_{3S} = (121.3\pm1.2)\times 10^{6}$ $\Upsilon(3S)$
and $N_{2S} = (98.3\pm0.9)\times 10^6$ $\Upsilon(2S)$ mesons, respectively.

\subsection{Searches for $\Upsilon(2S,3S) \rightarrow \gamma A^0 (\rightarrow hadrons)$}

The searches of $A^0$ use hadronic events in which the full event energy is reconstructed. In the events with at 
least two charged tracks, the highest energy photon is assumed to be the photon from the radiative $\Upsilon(nS)$ 
decay. The $A^0$ is reconstructed by adding the 4-momenta of the remaining particles, including $K^0_S\rightarrow\pi^+\pi^-$, 
$K^{\pm}$, $\pi^{\pm}$, $\pi^0\rightarrow\gamma\gamma$, and any unused photon. The radiative photon energy must be 
greater than $2.5~{\rm GeV}$ ($\Upsilon(3S)$) or $2.2~{\rm GeV}$ ($\Upsilon(2S)$). The $A^0$ mass resolution is 
improved by constraining the radiative photon and the $A^0$ decay products to come from the same vertex and the 
total 4-momentum to be that of the CM system.

\begin{figure}[hb!]
\begin{center}
\includegraphics[width=6.0cm,keepaspectratio]{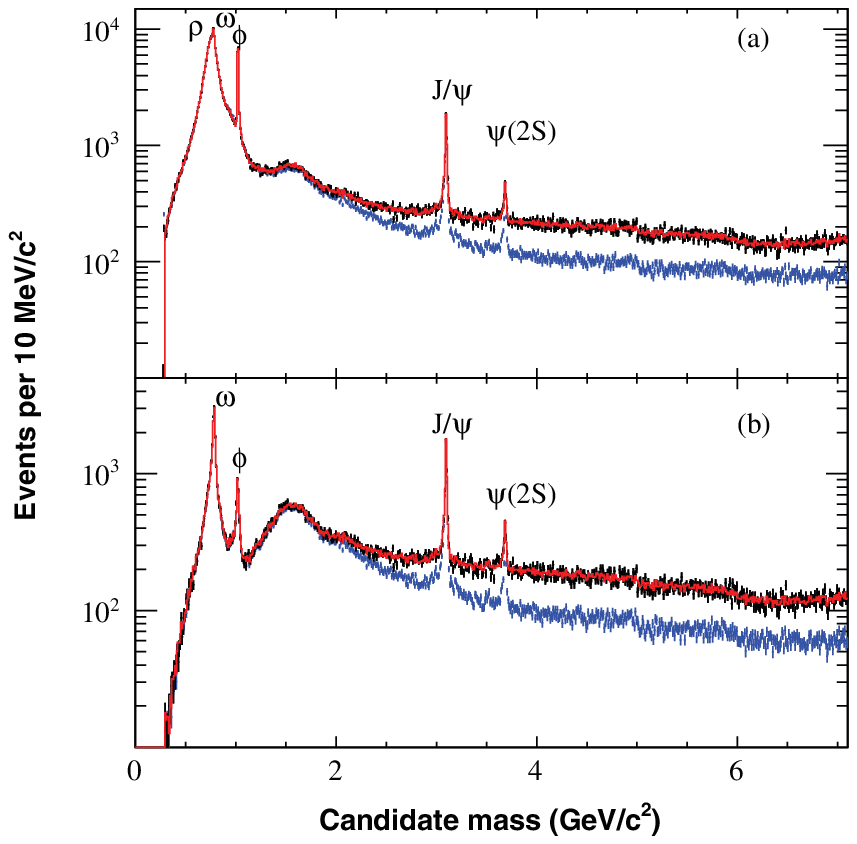}
\includegraphics[width=6.0cm,keepaspectratio]{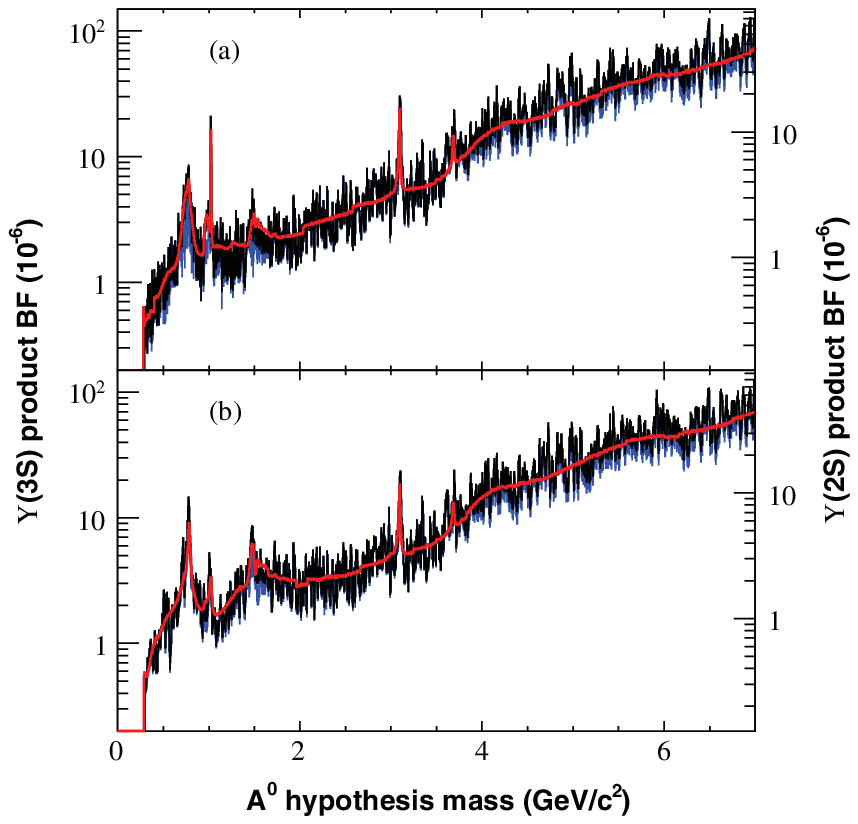}
\end{center}
\caption
{\label{fig:Upsilon2S3STogammaA0toHD}
{\em Left: Candidate mass spectrum in the (a) CP-all and (b) CP-odd analyses. Black dots are 
on-peak data, blue dots are off-peak scaled data and the red curve is the background fit. The prominent 
initial-state radiation resonances are labeled.
Right: 90\% C.L. upper limits on the product of $B(\Upsilon(3S)\rightarrow\gamma A^0)B(A^0\rightarrow hadrons)$ (left axis) 
and $B(\Upsilon(3S)\rightarrow\gamma A^0)B(A^0\rightarrow hadrons)$ (right axis), for (a) CP-all and (b) CP-odd analysis.}}
\end{figure}

Two parallel analysis are performed: one in which no assumption is made about the CP nature of the $A^0$, "CP-all";
and one in which the $A^0$ is assumed to be CP-odd ($A^0\rightarrow\pi^+\pi^-,K^+K^-$ are excluded). The analysis 
selects 371740 (171136) events for CP-all (CP-odd) in the combined $\Upsilon(2S,3S)$ dataset, with 
$0.29 < m_{A^0} < 7.1~{\rm GeV/c^2}$. The left hand plots of figure~\ref{fig:Upsilon2S3STogammaA0toHD} show 
the distributions of the reconstructed $A^0$ mass; a signal would appear as a narrow peak in these spectra. 
The number of signal events for a particular mass hypothesis of the $A^0$ is estimated as the number of events 
within the mass window (a bin in the spectrum) minus the number of background events. Since the $A^0$ is narrow, 
the mass window is chosen according to the $A^0$ mass resolution, which varies from $3$ to $26~{\rm MeV/c^2}$ as 
$m_{A^0}$ increases from $0.29$ to $7~{\rm GeV/c^2}$. The background events are estimated from continuum 
$\Upsilon(nS)$ events (continuum sample), {\it i.e.} off-peak $\Upsilon(2S,3S,4S)$ data and on-peak $\Upsilon(4S)$ 
data. The dominant background is due to $e^+e^-\rightarrow q\bar{q}$ (continuum), and consists mostly of initial-state 
radiation (ISR) production of light mesons and non-resonant hadrons.

The number of background events is obtained from a fit to the data sample aggregating the properly 
scaled datasets just mentioned. The $A^0$ signal is 
evaluated at masses from $0.291$ to $7.000~{\rm GeV/c^2}$ ($0.300$ to $7.000~{\rm GeV/c^2}$) 
in $1~{\rm MeV/c^2}$ steps for the CP-all (CP-odd) analysis. The largest upwards fluctuations are 
$2.8\sigma$ ($2.2\sigma$) at $3.107~{\rm GeV/c^2}$ ($0.772~{\rm GeV/c^2}$) for CP-all (CP-odd) analysis, 
meaning that no evidence of a signal is found. Therefore, upper limits on the product of branching fractions
$B(\Upsilon(nS)\rightarrow\gamma A^0)B(A^0\rightarrow hadrons)$ (right hand plot of 
figure~\ref{fig:Upsilon2S3STogammaA0toHD}) are derived, ranging from $1\times10^{-6}$ at 
$0.3~{\rm GeV/c^2}$ to $8\times10^{-5}$ at $7~{\rm GeV/c^2}$
at the 90\% C.L~\cite{BaBar_Upsilon2S3S_to_gammaA0_HD}.

\subsection{Searches for $\Upsilon(1S) \rightarrow \gamma A^0 (\rightarrow \mu^+\mu^-)$}

These searches look for a di-muon resonance in the fully reconstructed decay chains 
$\Upsilon(2S,3S)\rightarrow\pi^+\pi^-\Upsilon(1S)(\rightarrow \gamma A^0)$, 
$A^0\rightarrow\mu^+\mu^-$. Events are selected which contain exactly four charged tracks 
and a single photon with CM energy larger than $200~{\rm MeV}$. All the tracks must be 
consistent with originating from the collision point, and at least one must be identified 
as a muon. The two tracks with the highest momenta are assumed to be muons and are constrained 
to originate from a common $A^0$ vertex. The $\Upsilon(1S)$ candidate is reconstructed by 
combining the $A^0$ candidate and the photon. Finally, the $\Upsilon(2S,3S)$ are formed with 
the $\Upsilon(1S)$ candidate with the two remaining opposite charged tracks, assumed to be 
pions. The resolution in the di-muon mass is improved by performing a kinematical fit of the 
full decay chain, where the $\Upsilon(2S,3S)$ vertex is constrained to the collision point, 
 mass constraints of the $\Upsilon(2S,3S)$ and $\Upsilon(1S)$ candidates are enforced, 
 as well as the requirement that the $\Upsilon(2S,3S)$ energy be consistent with the $e^+e^-$ 
 CM energy.

\begin{figure}[hb!]
\begin{center}
\includegraphics[width=6.0cm,keepaspectratio]{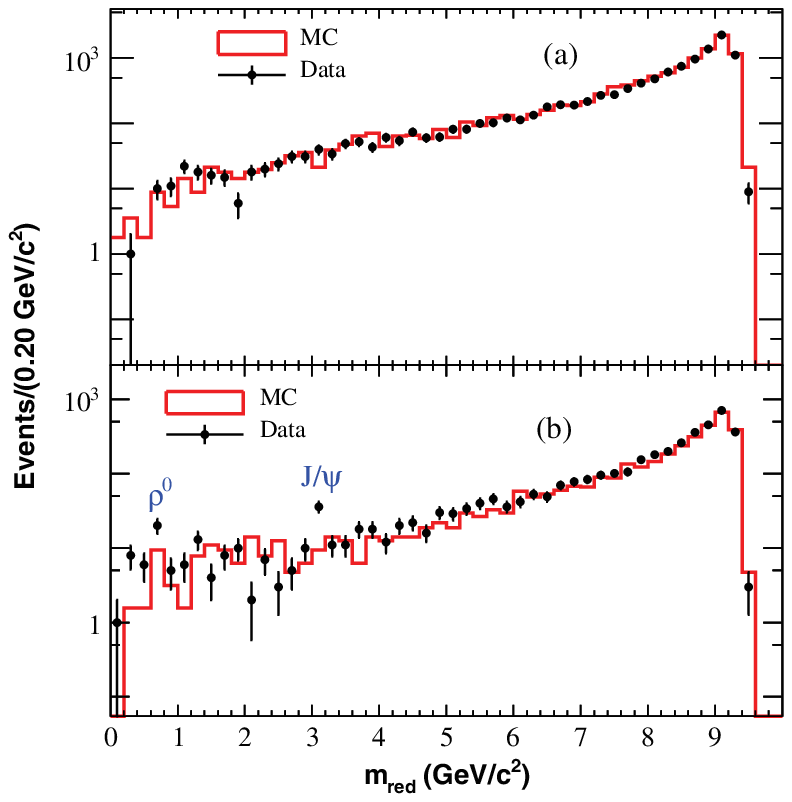}
\includegraphics[width=4.0cm,keepaspectratio]{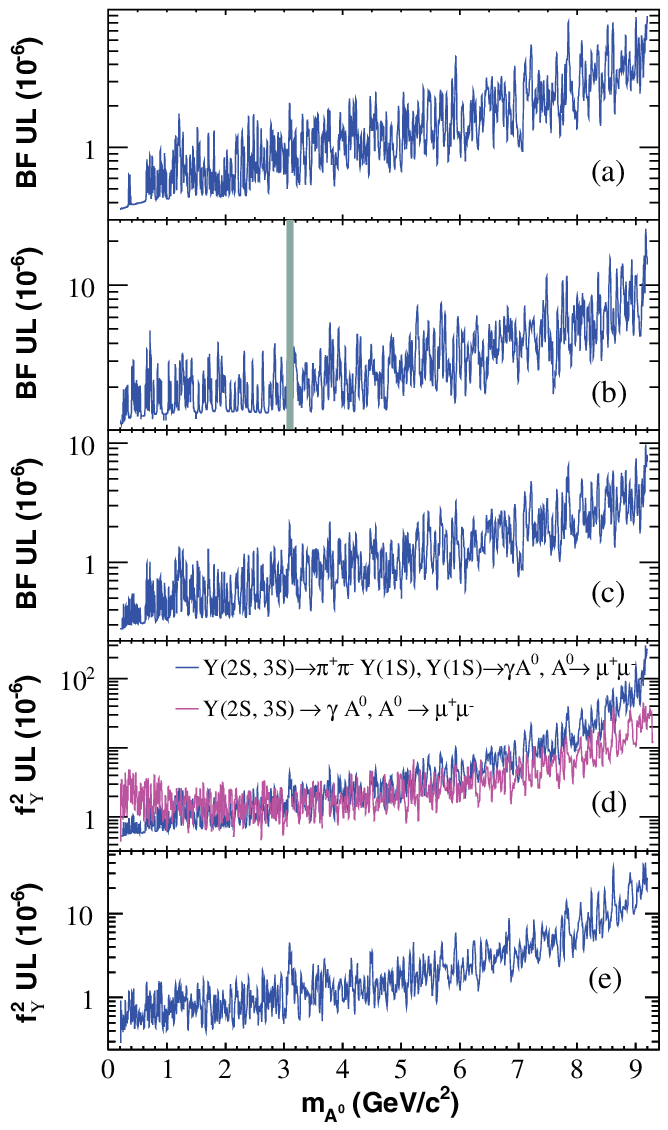}
\end{center}
\caption
{\label{fig:Upsilon1STogammaA0tomumu}
{\em Left: $m_{\rm red}$ distribution for (a) $\Upsilon(2S)$ and (b) $\Upsilon(3S)$ datasets. Peaking background 
components can be seen for $\Upsilon(3S)$ data.
Right: 90\% upper limits on the product of branching fractions $B(\Upsilon(1S)\rightarrow\gamma A^0)B(A^0\rightarrow \mu^+\mu^-)$ 
for (a) $\Upsilon(2S)$, (b) $\Upsilon(3S)$ and (c) combined $\Upsilon(2S,3S)$ datasets; (d) 90\% upper limit 
on $f^2_{\Upsilon}\times B(A^0\rightarrow \mu^+\mu^-)$ (blue curve), together with previous {\babar} 
measurements (magenta curve), and (e) combined limit.}}
\end{figure}

A total of 11136 $\Upsilon(2S)$ (3857 $\Upsilon(3S)$) candidates are selected. A resonant 
peak from the $A^0$ decay is expected in the di-muon reduced mass spectrum 
($m_{\rm red} = \sqrt{m^2_{\mu^+\mu^-} - 4m^2_{\mu}}$). The distribution of $m_{\rm red}$ 
for the selected candidates is shown on the left hand plots of figure~\ref{fig:Upsilon1STogammaA0tomumu}.
The main backgrounds are dominated by non-resonant di-muon decays. The signal yield is extracted as a 
function of $m_{A^0}$ in the region $0.212 \le m_{A^0} \le 9.20~{\rm GeV/c^2}$ by performing a series 
of one-dimensional extended maximum likelihood fits to the $m_{\rm red}$ distribution. The $A^0$ signal 
is searched in steps of half the $m_{\rm red}$ resolution, resulting in a total of 4585 points. 
The largest upward fluctuations are found to be $3.62\sigma$ ($2.97\sigma$) at $m_{A^0} = 7.87~{\rm GeV/c^2}$ 
($3.78~{\rm GeV/c^2}$) for $\Upsilon(2S)$ ($\Upsilon(3S)$) dataset. The probabilities to observe such 
fluctuation are estimated to be 18.1\% (66.2\%). Therefore, the distribution of the signal significance 
is compatible with the null hypothesis.

We find no significant signal and set 90\% C.L. Bayesian upper limits on the 
product of branching fractions $B(\Upsilon(1S)\rightarrow\gamma A^0)B(A^0\rightarrow\mu^+\mu^-)$ 
in the range of $0.212 \le m_{A^0} \le 9.20~{\rm GeV/c^2}$ (right hand plot of 
figure~\ref{fig:Upsilon1STogammaA0tomumu}). The limits range between $(0.37 - 8.97)\times10^{-6}$ for 
$\Upsilon(2S)$ data, $(1.13 - 24.2)\times10^{-6}$ for $\Upsilon(3S)$ data and $(0.28 - 9.7)\times10^{-6}$ 
for the combined $\Upsilon(2S,3S)$ datasets~\cite{BaBar_Upsilon1S_to_gammaA0_mumu}. By using the relation
$B(\Upsilon(nS)\rightarrow\gamma A^0)/B(\Upsilon(nS)\rightarrow\ell^+\ell^-) = (f^2_{\Upsilon}/2\pi\alpha_e) (1 - m^2_{A^0}/m^2_{\Upsilon(nS)})$, 
where, $n=1,2,3$, $\ell = e, \mu$, $\alpha_e$ is the fine structure constant and $f_{\Upsilon}$ the 
effective Yukawa coupling~\cite{BaBar_Upsilon1S_to_gammaA0_mumu} of the b-quark to $A^0$, we 
set 90\% upper limits on the product $f^2_{\Upsilon}\times B(A^0\rightarrow\mu^+\mu^-)$ using the 
results from the combined $\Upsilon(2S,3S)$ datasets. The upper limit ranges from $0.54\times10^{-6}$ to $3.0\times10^{-4}$ 
depending upon the $A^0$ mass. Combining the present results with the previous {\babar} results~\cite{BaBar_previous} 
on $\Upsilon(2S,3S)\rightarrow\gamma A^0$ we obtain a 90\% upper limit on $f^2_{\Upsilon}\times B(A^0\rightarrow\mu^+\mu^-)$
in the range $(0.29-40)\times10^{-6}$ for $m_{A^0} \le 9.2~{\rm GeV/c^2}$ (right hand plot of 
figure~\ref{fig:Upsilon1STogammaA0tomumu}).

\subsection{Searches for $\Upsilon(1S) \rightarrow \gamma A^0 (\rightarrow \tau^+\tau^-)$}

For this search, the $\Upsilon(2S) \rightarrow \pi^+\pi^- \Upsilon(1S)$ transition is used.
A signal candidate consist of a photon plus four charged tracks, two associated with the pions of 
the transition between $\Upsilon(2S)$ and $\Upsilon(1S)$ and two from one-prong decays of each 
$\tau$-lepton ($\tau^+\rightarrow e^+\bar{\nu}_{e}\nu_{\tau}$,$\mu^+\bar{\nu}_{\mu}\nu_{\tau}$,
$\pi^+\nu_{\tau}$). We require that at least one $\tau$ decays leptonically, and examine 
five combinations of daughters: $ee$, $e\mu$, $\pi\mu$, $\mu\mu$, $\mu\pi$.
The masses of the $\Upsilon(1S)$ and $A^0$ candidates are calculated from the 
$m^2_{\rm recoil} = M^2_{\Upsilon(2S,3S)} + m^2_{\pi\pi} - 2M_{\Upsilon(2S,3S)}E^{CM}_{\pi\pi}$
and 
$m^2_X = (P_{e^+e^-} - P_{\pi\pi} - P_{\gamma})^2$ kinematic variables, where the $P$ are the 
four-momenta of the indicated system. 

\begin{figure}[hb!]
\begin{center}
\includegraphics[width=6.0cm,keepaspectratio]{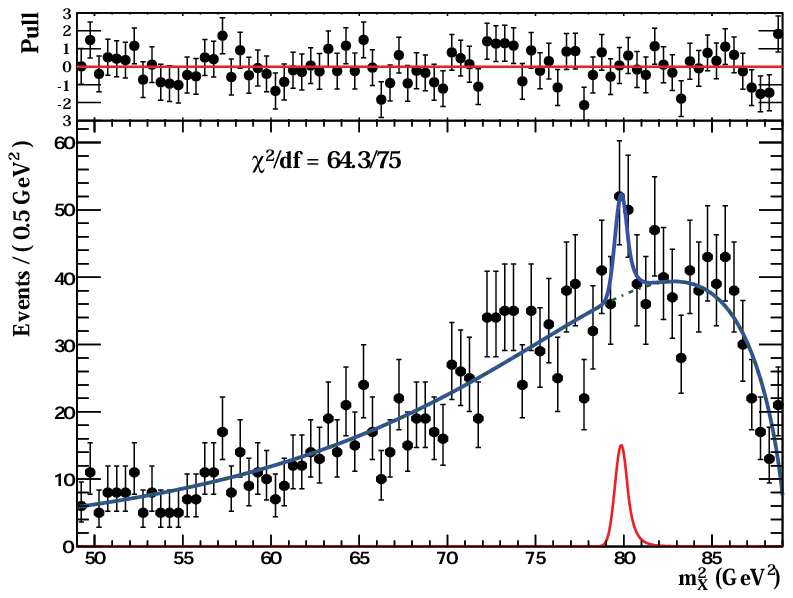}
\includegraphics[width=6.0cm,keepaspectratio]{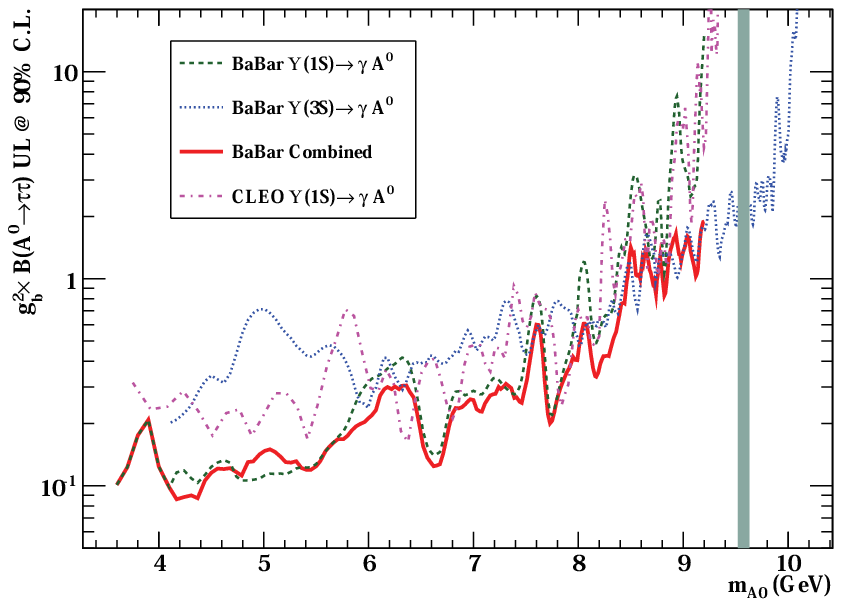}
\end{center}
\caption
{\label{fig:Upsilon1STogammaA0totautau}
{\em Left: Fit to $m^2_X$ distribution for the $m_{A^0}$ point that return the 
largest upward fluctuation (see text). Right: 90\% C.L. upper limits for $g^2_b\times B(A^0\rightarrow \tau^+\tau^-)$. 
Shown are the present results (dashed green), the previous {\babar} results from $\Upsilon(3S)$ radiative decays 
(dotted blue), the combination (red); and results from CLEO experiment (dashed magenta).}}
\end{figure}

A resonant peak from the $A^0$ decay would manifest in the $m^2_X$ spectrum. We extract 
the signal yields as a function of the $m_{A^0}$ in the interval
$3.6 \le m_{A^0} \le 9.2~{\rm GeV/c^2}$ by performing a series of maximum-likelihood fits 
to $m^2_X$. The fit model contains contributions from signal, which is expected to peak near 
the light Higgs mass squared, and a smooth background function, arising from continuum and 
radiative leptonic $\Upsilon(1S)$ backgrounds. We search for the $A^0$ in varying mass steps 
that correspond to approximately half of the expected resolution on $m_{A^0}$. A total of 201 
mass points are sampled. The most significant upward fluctuations occurs with $2.7\sigma$ at 
$m_{A^0} = 6.36~{\rm GeV/c^2}$ (left hand plot of figure~\ref{fig:Upsilon1STogammaA0totautau}). 
Estimations reveal that such fluctuations are expected with a 
probability of 7.5\%, therefore, we conclude that no significant $A^0$ signal is found and 
set Bayesian 90\% C.L. upper limits on the product 
$B(\Upsilon(1S)\rightarrow\gamma A^0)B(A^0\rightarrow\tau^+\tau^-)$, computed with a uniform prior.
The limits range between $(0.9 - 13.0)\times10^{-5}$~\cite{BaBar_Upsilon1S_to_gammaA0_tautau}.
Using the relation 
$B(\Upsilon(nS)\rightarrow\gamma A^0)/B(\Upsilon(nS)\rightarrow\ell^+\ell^-) = (g^2_b G_F m^2_b/\sqrt{2}\pi\alpha_e) {\mathcal F}_{QCD} (1 - m^2_{A^0}/m^2_{\Upsilon(nS)})$, 
where $g^2_b$ is the Yukawa coupling of the b-quark to the $A^0$, $G_F$ the Fermi constant, and ${\mathcal F}_{QCD}$ 
includes QCD corrections~\cite{BaBar_Upsilon1S_to_gammaA0_tautau}; we can set a constrain on the product $g^2_b\times B(A^0\rightarrow\tau^+\tau^-)$, and 
combine the present results with previous {\babar} measurements on 
$B(\Upsilon(3S)\rightarrow\gamma A^0)\times B(A^0\rightarrow\tau^+\tau^-)$~\cite{BaBar_previous}. 
We set a 90\% C.L. upper limits on the product $g^2_b\times B(A^0\rightarrow \tau^+\tau^-)$ in the 
range $0.09 - 1.9$ for $m_{A^0} \le 9.2~{\rm GeV/c^2}$ (right hand plot of figure~\ref{fig:Upsilon1STogammaA0totautau}). 
Our limits place significant constraints on NMSSM parameter space.

\newpage
\section{Dark Higgs and dark photon searches at {\babar}}

There is overwhelming evidence for dark matter from terrestrial and satellite astrophysical observations, 
but its precise nature and origin remain unknown. To try to explain this observational evidence many models
~\cite{Dark_matter_models} introduce a new hidden dark sector under which WIMP-like dark matter particles
are charged. In these models, the WIMP-like particle can annihilate into pairs of dark bosons, which subsequently
annihilate into leptons (protons are kinematically forbidden). In one of these models~\cite{Dark_matter_models} there is a new dark sector 
that couples to the SM particles with a dark photon, $A'$, through a small kinetic term. Astrophysical 
data constrains the $A'$ mass to be a few ${\rm GeV}$. The $A'$ acquire its mass via the Higgs mechanism, 
adding a dark Higgs, $h'$, to the theory. The mass hierarchy is not constrained, and the $h'$ could be 
light as well. The high collision rate and the relatively clean experimental environment of B-factories 
is an ideal place to probe for ${\rm MeV-GeV}$ dark matter, complementing the searches performed at LHC.

\subsection{Searches for $e^+e^- \rightarrow A'h'(\rightarrow A'A')$; with $A'\rightarrow e^+e^-,~\mu^+\mu^-,~\pi^+\pi^-$}

We search for dark Higgs and dark photons via {\it Higgsstrahlung} production 
$e^+e^-\rightarrow A'^*\rightarrow A'h'(\rightarrow A'A')$~\cite{DarkPhoton_BaBar}. This interaction 
is very interesting because it is singly suppressed by the mixing strength between the SM and the dark 
sector, $\epsilon$. If observed, it could provide an unambiguous signature of NP. The event topology depends on the 
$A'$ and $h'$ masses. This measurement is performed in the range $0.8 < m_{h'} < 10.0~{\rm GeV/c^3}$ 
and $0.25 < m_{A'} < 3.0~{\rm GeV/c^3}$ with the constrain $m_{h'} > 2 m_{A'}$. The data sample used 
consists of $521~{\rm fb}^{-1}$ collected mainly at the $\Upsilon(4S)$ peak, but also includes data 
collected at the  $\Upsilon(2S)$ and $\Upsilon(3S)$ peaks, as well as off-peak data.

\begin{figure}[hb!]
\begin{center}
\includegraphics[width=6.0cm,keepaspectratio]{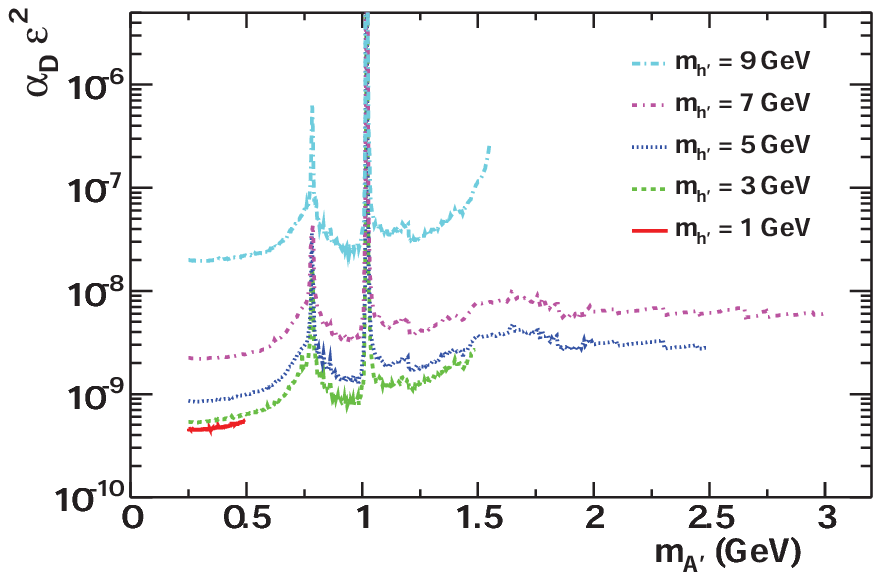}
\includegraphics[width=6.0cm,keepaspectratio]{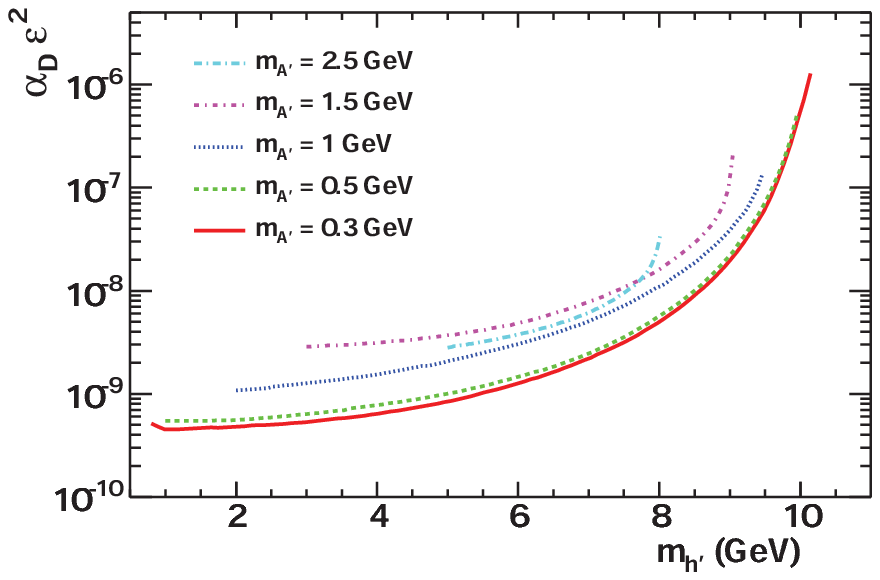}
\end{center}
\caption
{\label{fig:DarkPhoton}
{\em 90\% C.L. upper limits on $\alpha_D\epsilon^2$ (see text). Left: limits as a function of $m_{A'}$ 
for selected values of $m_{h'}$; Right: limits as a function of $m_{h'}$ for selected values of $m_{A'}$.}}
\end{figure}

The process $e^+e^-\rightarrow A'^*\rightarrow A'h'(\rightarrow A'A')$ is either fully reconstructed 
in $3(\ell^+\ell^-)$, $2(\ell^+\ell^-)\pi^+\pi^-$ and $\ell^+\ell^-2(\pi^+\pi^-)$ ($\ell = e,\mu$); 
or partially reconstructed in $2(\mu^+\mu^-) + X$ and $(\mu^+\mu^-)(e^+e^-) + X$, where $X$ denotes 
any final state other than a pair of leptons or pions. A total of six events where selected: 
one $4\mu2\pi$, two $2\mu4\pi$, two $2e4\pi$ and one $4\mu + X$. No candidate with six leptons survives 
the selection. The selected events are very likely to come from $\rho\rightarrow\pi^+\pi^-$ or 
$\omega\rightarrow\pi^+\pi^-$ decays near $m_{A'} \sim 0.7 - 0.8~{\rm GeV/c^2}$, and it is concluded 
that no significant signal is observed.

Using a Bayesian method, upper limits on the $e^+e^-\rightarrow A'^*\rightarrow A'h'(\rightarrow A'A')$ 
cross-section are obtained as a function of $m_{h'}$ and $m_{A'}$. These limits can be translated into 
90\% upper limit on the coupling $\alpha_D\epsilon^2$ (see figure~\ref{fig:DarkPhoton}), 
where $\alpha_D = g^2_D/4\pi$, with $g_D$ the 
dark sector gauge coupling. Values as low as $10^{-10} - 10^{-8}$ are excluded for a large range of 
$m_{A'}$ and $m_{h'}$, assuming prompt decay. Under the assumption that $\alpha_D \approx \alpha_e$, 
the current measurement can also be translated into limits on the mixing strength $\epsilon$ which 
range $10^{-4} - 10^{-3}$, an order of magnitude smaller that current experimental bounds from direct 
photon production in this mass range.

\section{Conclusion}
We performed several searches for evidence of dark sector candidates and CP-odd light Higgs
in the $\Upsilon(2S,3S,4S)$ {\babar} data sample. The data show no evidence of such signals but
enable to improve the limits within the parameter space of various NP models.

\end{document}